\documentclass[runningheads]{llncs}
\usepackage[T1]{fontenc}
\usepackage{multirow} 
\usepackage{graphicx}
\usepackage{makecell}
\usepackage[table]{xcolor}
\usepackage{amsmath,amssymb,amsfonts}

\newcommand{\sone}{\colorbox{yellow!25}{S1}}
\newcommand{\stwo}{\colorbox{orange!25}{S2}}
\newcommand{\sthree}{\colorbox{red!25}{S3}}

\newcommand{\ethree}{\colorbox{orange!25}{E3}}
\newcommand{\efour}{\colorbox{red!25}{E4}}

\newcommand{\ctwo}{\colorbox{orange!25}{C2}}
\newcommand{\cthree}{\colorbox{red!25}{C3}}

\newcommand{\asila}{\colorbox{green!25}{A}}
\newcommand{\asilb}{\colorbox{yellow!25}{B}}
\newcommand{\asilc}{\colorbox{orange!25}{C}}
\newcommand{\asild}{\colorbox{red!25}{D}}

\begin{document}
\title{Towards a Systematic Risk Assessment of Deep Neural Network Limitations in Autonomous Driving Perception}
\titlerunning{Towards a Systematic Risk Assessment of DNN Limitations in AD}
%
\author{Svetlana Pavlitska\inst{1,2} \and
Christopher Gerking\inst{2} \and
J. Marius Zöllner\inst{1,2}}
\authorrunning{S. Pavlitska et al.}
\institute{FZI Research Center for Information Technology, Germany \and
Karlsruhe Institute of Technology, Germany\\
\email{pavlitska@fzi.de}\\
}

\maketitle            
\begin{abstract}
Safety and security are essential for the admission and acceptance of automated and autonomous vehicles. Deep neural networks (DNNs) are widely used for perception and further components of the autonomous driving (AD) stack. However, they possess several limitations, including lack of generalization, efficiency, explainability, plausibility, and robustness. These insufficiencies can pose significant risks to autonomous driving systems. However, hazards, threats, and risks associated with DNN limitations in this domain have not been systematically studied so far. In this work, we propose a joint workflow for risk assessment combining the hazard analysis and risk assessment (HARA) following ISO 26262 and threat analysis and risk assessment (TARA) following the ISO/SAE 21434 to identify and analyze risks arising from inherent DNN limitations in AD perception.

\end{abstract}

\section{Introduction}

In autonomous driving (AD), deep neural networks (DNNs) are a significant source of perception capability and a potential source of malfunctions due to their limitations, such as lack of generalization, transparency, or robustness~\cite{houben2022inspect}. Risks posed by these limitations must be systematically studied because they directly impact perception reliability, safety, and trustworthiness, which are critical for ensuring safe vehicle behavior in complex and unpredictable environments.

The established risk assessment frameworks for AD, such as HARA (Hazard Analysis and Risk Assessment)~\cite{iso26262} and TARA (Threat Analysis and Risk Assessment)~\cite{iso21434}, and emerging standards addressing artificial intelligence (AI) malfunctions in general~\cite{iso24028} and in AD~\cite{iso8800} usually address all possible malfunctions in an autonomous vehicle (AV), not focusing on DNN-rooted ones. Furthermore, older standards focus on road vehicles of SAE levels 1-2 of driving automation~\cite{saeJ3016}. DNNs, however, play a bigger role in AVs of levels 3-4. DNN-rooted errors can, therefore, have a bigger impact on the safety of these vehicles.

There remains a \textbf{research gap} in systematically addressing risks that stem specifically from DNN limitations. While risks arising from some limitations, including the lack of robustness and explainability, have been considered more often, others, like the lack of plausibility, have received less attention in systematic risk analysis. Moreover, DNN limitations cause both safety and security risks, leading to failure modes that are not fully addressed by safety or security analysis alone. 
Our work aims to close this gap by performing risk analysis while focusing on both safety and security implications of DNN-based perception systems in AVs of SAE levels 3-4. This work's key \textbf{contributions} are: 
\begin{itemize}
    \item Analysis of the treatment of DNN limitations in existing AD and AI safety and security standards, highlighting the need for a unified risk assessment;
    \item Proposal of a HARA-TARA workflow tailored to DNN-based perception;
    \item Demonstration of the workflow on five key DNN limitations (lack of generalization, efficiency, explainability, plausibility, and robustness) and a discussion of safety-security interdependencies.
\end{itemize}

\section{Background and Related Work}

\subsection{DNN Limitations}
DNNs exhibit limitations arising from either their nature, i.e., intrinsic architectural properties such as linear decision boundaries or lack of causal reasoning, or from nurture, i.e., external factors like biased data, insufficient training, or poor labeling. While nurture-related issues can often be addressed through better data or training, nature-based limitations are structurally inherent and more difficult to resolve. For instance, adversarial vulnerability may not simply result from poor training but from the intrinsic fragility of high-dimensional linear decision boundaries~\cite{ilyas2019adversarial}.  Similarly, the lack of explainability may either result from nurture-based factors such as poor documentation or inadequate tooling, or reflect the fundamentally statistical and non-causal nature of DNNs.

We use the terms \textbf{DNN limitation} and \textbf{DNN insufficiency} as synonyms and rely on the following definition: \textit{"DNN insufficiency is a property of a trained DNN model that has a negative impact for the use in safety-relevant systems, and is inherent to the technology of DNNs. A measurable DNN insufficiency causes errors that decrease associated performance measures"}~\cite{schwalbe2020structuring}. 

There is no common DNN limitations taxonomy, although several have been proposed~\cite{willers2020safety,schwalbe2020structuring,saemann2020strategy}. ISO PAS 8800~\cite{iso8800} groups safety-related properties of AI systems by the scope: model-related properties include AI robustness, generalization capability, reliability, predictability, bias, and fairness. This list is not tailored for perception in AVs and, therefore, too broad for our case. ISO-IEC TR 5469~\cite{iso5469} defines six properties for the automotive camera perception: specificability, generalization, domain shift, robustness-safeness, diversity, and confidence. This list captures high-level system concerns but is less directly tied to DNN technical behavior. We use the taxonomy by Sämann et al.~\cite{saemann2020strategy} for perception safety in highly automated driving, which captures DNN-specific limitations directly translating into safety and cybersecurity risks for AD:

\textbf{1. Lack of generalization}: DNNs often fail to generalize beyond the training data and struggle to perform reliably on data that differs from the training distribution, such as in rare, novel, or adversarial conditions. For AD, long-tail driving scenarios, corner cases, and unseen events present a problem~\cite{bogdoll2022anomaly}.

\textbf{2. Lack of efficiency}: Due to deep architectures and a large number of parameters, DNNs demand significant hardware resources, making the deployment in resource-constrained environments like AVs challenging. Optimizations such as pruning~\cite{han2015deep} typically come at the cost of accuracy or robustness.

\textbf{3. Lack of explainability\footnote{We use the term “explainability” in line with Sämann et al.~\cite{saemann2020strategy} and standards such as ISO PAS 8800, which reference explainability in the context of AI-based systems. In broader AI literature, however, a distinction is often made between “explainability” referring to post-hoc explanations and “interpretability” referring to a model’s inherent transparency~\cite{rudin2019stop}.}}:  DNNs have complex, non-transparent internal representations and decision-making processes, which limit the ability of developers and users to understand, trust, and verify their outputs.

\textbf{4. Lack of plausibility}: DNNs can assign high confidence to outputs that violate physical laws or common-sense constraints. Examples for AD are pedestrians or cars floating mid-air or having unrealistic shapes ~\cite{adilova2021plants,vivekanandan2022plausibility}.

\textbf{5. Lack of robustness}: DNNs demonstrate reduced performance for input perturbations and, in particular, are vulnerable to purposely generated \textbf{adversarial attacks}~\cite{szegedy2013intriguing,goodfellow2014explaining}. Adversarial attacks can also be performed in the form of an \textbf{adversarial patch}~\cite{brown2017adversarial}; adversarial patch attacks on different perception systems in AD have already been demonstrated~\cite{eykholt2018robust,nesti2022evaluating,pavlitska2025fool,pavlitskaya2020feasibility,wei2023adversarial}. 

\subsection{Standards and Norms for AD and AI Safety and Security}
\label{sec:standards}

The safety of AI components in AVs has been addressed in several standards and norms (see Table~\ref{tab:standards_coverage}). Neither ISO 26262~\cite{iso26262} nor ISO/SAE 21448~\cite{iso21448} explicitly mentions DNNs. ISO PAS 8800~\cite{iso8800} provides the most comprehensive guidance for machine learning (ML)-specific risks, directly addressing DNN shortcomings like lack of robustness, generalization, and explainability, and recommending mitigation measures such as redundancy, robustness testing, and data augmentation. While not an ISO standard, ANSI/UL 4600~\cite{ul4600} explicitly identifies DNN-related challenges, including adversarial vulnerabilities and non-determinism, and requires rigorous justification of safety claims. Though more general, ISO/IEC TR 24028~\cite{iso24028} and TR 24029~\cite{iso24029} partially cover DNN limitations by emphasizing robustness, uncertainty estimation, and performance validation. While high-level, the EU AI Act~\cite{euact},  supports DNN risk mitigation by mandating transparency, robustness, and human oversight in high-risk AI applications such as AD. In summary, due to the heterogeneity of DNN limitations, these standards do not fully cover them. 

\begin{table*}[t]
\centering
\caption{Coverage of DNN limitations in relevant automotive, safety, security, and AI standards. The following abbreviations for DNN limitations are used: G - generalization, E - efficiency, X - explainability, P - plausibility, R - robustness.}
\label{tab:standards_coverage}
\resizebox{1.0\linewidth}{!}{ 
    \begin{tabular}{|l r|p{6cm}|ccccc|}
    \hline
    \textbf{Standard} & \textbf{Year} & \textbf{Description} & \textbf{G} & \textbf{E} & \textbf{X} & \textbf{P} & \textbf{R} \\
    \hline
    \textbf{ISO 26262}~\cite{iso26262} & 2018  & Functional safety of automotive electric/electronic systems, focus on system-level hazards, introduces HARA. &  &  &  &  &  \\
    \hline
    \textbf{ISO/IEC TR 24028}~\cite{iso24028} & 2020 & Trustworthiness in AI; high-level properties including reliability, robustness, and security. & $\checkmark$ &  & $\checkmark$ &  & $\checkmark$ \\
    \hline
    \textbf{ISO/SAE 21434}~\cite{iso21434} & 2021 & Automotive cybersecurity, introduces TARA. &  &  &  &  & $\checkmark$ \\
    \hline
    \textbf{ISO 21448}~\cite{iso21448} & 2022 & Safety of the Intended Functionality (SOTIF); covers non-fault-based hazards including sensor/ML insufficiencies. & $\checkmark$ &  &  & $\checkmark$ & $\checkmark$ \\
    \hline
    \textbf{ISO PAS 8800}~\cite{iso8800} & 2022  & Safety for ML components in automotive systems. & $\checkmark$ & $\checkmark$ &  &  & $\checkmark$ \\
    \hline
    \textbf{ANSI/UL 4600}~\cite{ul4600} & 2022 & AV safety standard including ML, uncertainty, and assurance methods. & $\checkmark$ & $\checkmark$ & $\checkmark$ & $\checkmark$ & $\checkmark$ \\
    \hline
    \textbf{ISO/IEC TR 24029}~\cite{iso24029} & 2022& Assessment of the robustness of DNNs; evaluation methodologies for AI reliability. &  &  &  &  & $\checkmark$ \\
    \hline
    \textbf{ISO/IEC TR 5469}~\cite{iso5469} & 2024 &  Guidance for risk management of AI systems; considers intended use, lifecycle, and trustworthiness. & $\checkmark$ &  & $\checkmark$ &  & $\checkmark$ \\
    \hline
    \textbf{EU AI Act}~\cite{euact} & 2024 & Regulatory framework for trustworthy AI; includes robustness, transparency, and risk-based classification. & $\checkmark$ &  & $\checkmark$ &  & $\checkmark$ \\
    \hline
    \end{tabular} 
}
\end{table*}

\subsection{Risk Analysis in AD}

Since HARA has been proposed in ISO 26262~\cite{iso26262}, its modifications and applications to various parts of the AD stack have been proposed. Beckers et al.~\cite{beckers2013astructured} suggested an ISO 26262-compliant structured and model-based HARA approach for automotive systems and demonstrated it on an example of an electronic steering column lock (ESCL) system. HARA was extended using fault-type guide words and an organized set of situations to determine relevant causes. Also, Object Constraint Language (OCL) validation checks were introduced. Warg et al.~\cite{warg2020quantitave} described how HARA can be applied to automated driving systems. Chia et al.~\cite{chia2021real} proposed an approach complementing HARA and failure modes and effects analysis (FMEA) for AV operations. We refer to a comprehensive survey~\cite{chia2022risk}, which provides an overview of risk assessment methods in AD, while not explicitly focusing on  AI-related risks. 

Risks stemming from adversarial attacks have also recently gained attention. Ghost et al.~\cite{ghosh2023anintegrated} applied TARA to AV perception and especially considered adversarial attacks on traffic signs as one of the scenarios. Grosse et al.~\cite{grosse2024qualitative} presented a qualitative analysis of AI security risks, measuring attack exposure and severity for state-of-the-art adversarial attacks on various tasks in the AD stack.  However, the authors have not perform the complete TARA.

\newpage
\clearpage
\section{Methodology}

\begin{figure}[h]
\includegraphics[width=\textwidth]{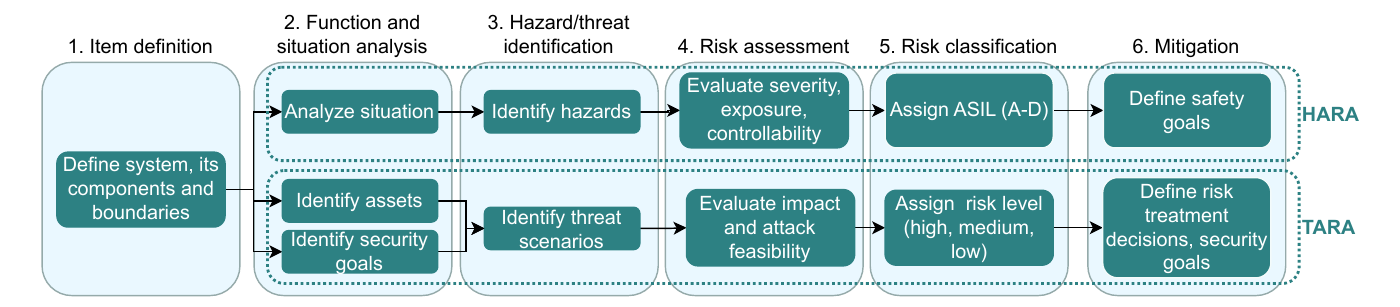}
\caption{Proposed combined HARA-TARA risk assessment.} 
\label{fig:process}
\end{figure}

We propose a workflow that leverages TARA and HARA steps (see Fig.~\ref{fig:process}). We assume at least an SAE Level 3~\cite{saeJ3016} AV with a standard modular pipeline, comprising sensing, perception, planning, and control. We primarily focus on the perception module of the pipeline. The workflow starts with system and context definition, then identifies hazards (sources of harm from a safety perspective) and threats (sources of harm from a security perspective). After that, risks are identified. HARA finalizes with a definition of safety goals, and TARA with decisions on risk treatment and security goals.
While HARA  and TARA individually provide structured methodologies for assessing safety and cybersecurity risks, respectively, their combination enables a holistic view that captures the interdependencies between safety hazards and security threats, critical when dealing with complex, opaque, and failure-prone AI components such as DNNs. By aligning the item definition and integrating situation analysis, threat/hazard identification, and risk evaluation, this workflow allows for the joint derivation and comparison of safety and security goals. 
We select HARA and TARA for the proposed workflow because they are mandated and explicitly structured within the key automotive standards.

\section{Evaluation}
We apply the proposed workflow to each DNN limitation, derive hazards/threats, and assign the Automotive Safety Integrity Level (ASIL)/risk levels.

\textbf{Step 1: Item Definition:} At the first stage, the \textbf{items} common for HARA and TARA are defined. An item in HARA is \textit{"a system \ldots{} or combination of systems \ldots{}, to which ISO 26262 is applied, that implements a function or part of a function at the vehicle level"} \cite{iso26262}.
In our use case, the \textbf{system} is a DNN-based perception module used for object detection, traffic sign recognition, semantic segmentation, and further driving scene understanding tasks. Inputs are camera images; the outputs are a list of detected objects or a semantic segmentation map for an input image; outputs are forwarded to the planning module.

\textbf{Step 2: Function and situation analysis:} The second stage aims to systematically identify item functions and examine how each insufficiency could result in functional degradation or failure. In HARA, the goal is to determine potential malfunctioning behaviors that could lead to a hazardous scenario. For our use case, the core function is accurately interpreting the driving environment across various environmental conditions. 

In the second TARA stage, \textbf{assets} are defined based on the item definition. An asset is anything of value that needs to be protected and whose compromise could lead to a damage scenario. We define six assets. Three are not directly impacted by DNN limitations: runtime environment, configuration and calibration data, model update and training pipeline. We consider only three remaining relevant assets: sensor data integrity, DNN behavior, and perception outputs. The second TARA stage also includes the identification of the \textbf{security goals}, i.e., high-level protection objectives derived from threats to identified assets. Also, high-level \textbf{security properties} (confidentiality, integrity, and availability) are mapped to specific security goals that protect identified assets from threats. We define the critical asset-property pairs and their corresponding security goals for the assets (see Table~\ref{tab:tara-assets}), establishing a foundation for the threat identification.

\begin{table}[t]
\centering
\caption{Assets, security properties, goals, and threat scenarios according to \textbf{TARA}.}
\label{tab:tara-assets}
\resizebox{1.0\linewidth}{!}{
    \begin{tabular}{|p{1.5cm}|l|p{1.7cm}|p{4cm}|p{7cm}|}
    \hline
    \textbf{Asset} & \textbf{Lim.} & \textbf{Security property} & \textbf{Security goal}  & \textbf{Threat scenario}\\
    \hline
    Sensor data integrity & R & Integrity & Prevent sensor data injection or modification. & Attacker places a crafted object in a scene, adding malicious noise to the sensor data.  \\
   \hline 
    DNN behavior  & G & Integrity & Maintain consistent performance across different driving scenarios. & Attacker deliberately puts a previously unseen object in front of a vehicle, causing a crash. \\
    & E & Integrity & Ensure timely inference for planning. &  Attacker injects rapid scene changes (e.g., flickering lights) to slow perception.\\
    & X & Non-repudiation & Enable traceability of model decisions. & Attacker causes stealthy misclassifications by exploiting the lack of decision traceability.\\
     
    & P & Integrity & Reject implausible inputs. & Attacker inserts objects of unrealistic shapes to exploit implausibility.\\
    & R & Integrity & Prevent model misbehavior under adversarial attack.  & Attacker uploads manipulated images into a training dataset, poisoning the DNN to behave incorrectly in specific contexts.\\
    
    \hline
    
    Perception outputs & G & Integrity & Prevent misinterpretation in long-tail scenarios & Attacker deliberately presents data unseen during training.\\
     & E & Availability & Ensure runtime reliability under hardware/software load & Attacker delays the transmission of perception outputs to the planning module. \\
      & P & Integrity & Enforce real-world consistency of outputs. & Attacker triggers implausible outputs (e.g., floating pedestrians), misleading planning.\\
    & R & Integrity & Avoid perception errors caused by adversarial input manipulations. & An adversarial sticker placed on a vehicle causes the DNN to interpret it as an empty road, failing to detect a collision threat.\\
      
     \hline
    \end{tabular}
}
\end{table}

\textbf{Steps 3: Hazard/Threat Scenario Identification:} Step 3 of the HARA and TARA identifies hazards and threat scenarios. In HARA, the focus is on identifying hazardous events resulting from functional failures. In TARA, the goal is to define threat scenarios that describe how an attacker could compromise the security properties of the identified assets. A threat scenario describes an attacker, a means of attack, the explored vulnerability, and the expected outcome. The found hazards and threats are used for further risk assessment.

\begin{table}[t]
\centering
\caption{Hazard identification and risk assessment according to \textbf{HARA (ISO 26262)}.}
\label{tab:hara_dnn}
\resizebox{1.0\linewidth}{!}{
\begin{tabular}{|l|p{4.5cm}|p{2cm}|p{2.5cm}|p{3cm}|l|p{5cm}|}
\hline
 \textbf{Lim.}  & \textbf{Hazard} & \textbf{Severity} & \textbf{Exposure} & \textbf{Controllability} & \textbf{ASIL} & \textbf{Safety Goal} \\
\hline
\textbf{G} & Failure to detect objects in unseen conditions leads to collision & 
\sthree{}: potential fatal injury via crash & 
\efour{}: likely in diverse driving conditions & 
\cthree{}: driver unable to intervene in SAE Level 4 and above & \asild{} & 
Ensure reliable detection and classification of objects in diverse and previously unseen driving scenarios \\
\hline

\hline
\textbf{E} & Delayed perception response causes failure to brake or steer in time & 
\sthree{}: collision possible due to delay & 
\ethree{}: common traffic exposure & 
\cthree{}: no timely driver override & \asilc{} & 
Ensure timely processing and delivery of perception information within safety-critical deadlines \\
\hline
\textbf{X} & Inability to validate or debug model hides failure modes during operation & 
\sone{}: indirect safety risk & 
\efour{}: affects every use case & 
\cthree{}: driver unaware of malfunction & \asilb{} & 
Ensure that perception failures can be monitored and traced to support safe system behavior \\
\hline
\textbf{P} & Implausible outputs (e.g., false obstacles) trigger incorrect maneuvers & 
\stwo{}: moderate injuries possible & 
\ethree{}: moderate likelihood & 
\ctwo{}: driver can possibly regain control & \asila{} & 
Ensure plausibility checks are applied to perception outputs to avoid incorrect vehicle behavior \\
 \hline
\textbf{R} & Misclassification from crafted adversarial inputs causes unsafe behavior (crash or missed detection) & 
\sthree{}: severe harm or fatalities via crash & 
\efour{}: feasible in open environment & 
\cthree{}: no fallback to driver in SAE Level 4 and above & \asild{} & 
Ensure robustness of perception models against adversarial manipulation in sensor inputs or the environment \\
\hline
\end{tabular}
}

\end{table}

\begin{table}[t]
\caption{Risk assessment according to \textbf{TARA (ISO/SAE 21434)}.}
\label{tab:tara_dnn}
\centering
\resizebox{1.0\linewidth}{!}{
\begin{tabular}{|l|p{2.5cm}|p{5cm}|p{2cm}|p{2.5cm}|l|p{4.2cm}|}
\hline
  \textbf{Lim.}  & \textbf{Affected } & \textbf{Threat Scenario} & \textbf{Impact} & \textbf{Feasibility} & \textbf{Risk} & \textbf{Risk Treatment Decision} \\
   & \textbf{asset} & &  & & \textbf{level} & \\
\hline

  \textbf{G} & DNN behavior & Unseen object causes model failure & \colorbox{red!30}{High}: crash hazard & \colorbox{orange!30}{Medium}: rare input & \colorbox{red!30}{High} & Risk reduction: data augmentation, OOD detection \\

   & Perception model outputs& Data shift leads to detection failure & \colorbox{red!30}{High}: detection missed & \colorbox{orange!30}{Medium}: rare input & \colorbox{red!30}{High} & Risk reduction: uncertainty estimation, retraining \\ \hline
  
   \textbf{E} & DNN behavior & Fast-changing scenes degrade perception timing & \colorbox{orange!30}{Medium}: late response & \colorbox{orange!30}{Medium}: hard to exploit & \colorbox{orange!30}{Medium} & Risk reduction: model acceleration, scheduling \\

    & Perception model outputs& Output latency affects planning & \colorbox{orange!30}{Medium}: delay risk & \colorbox{yellow!30}{Low}: internal issue & \colorbox{yellow!30}{Low} & Risk acceptance with monitoring \\ \hline

   \textbf{X} & DNN behavior & Attacker manipulates input to induce stealth misclassification, hidden by lack of traceability & \colorbox{orange!30}{Medium}: stealth risk & \colorbox{orange!30}{Medium}: traceability gap & \colorbox{orange!30}{Medium} & Risk reduction: XAI, logging \\ \hline

\textbf{P} & DNN behavior  &  Attacker injects a semantically inconsistent object to induce wrong control decisions & \colorbox{orange!30}{Medium}: delayed action & \colorbox{yellow!30}{Low}: detectable if monitored & \colorbox{orange!30}{Medium} & Risk reduction: plausibility constraints \\
\hline

\textbf{R} & Sensor data integrity & Attacker places a crafted object in a scene, adding malicious noise to sensor data & \colorbox{red!30}{High}: misleads detection & \colorbox{orange!30}{Medium}: physical access & \colorbox{red!30}{High} & Risk reduction: robust training, detection \\

   & DNN behavior & Attacker uploads manipulated images into training data, poisoning DNN behavior & \colorbox{red!30}{High}: stealth misbehavior & \colorbox{red!20}{High}: easy injection & \colorbox{red!30}{High} & Risk reduction: dataset integrity monitoring \\

   & Perception model outputs& Sticker on a vehicle fools the DNN to see an empty road & \colorbox{red!30}{High}: potential crash & \colorbox{orange!30}{Medium}: needs physical proximity & \colorbox{red!30}{High}& Risk reduction: physical robustness, adversarial training \\

\hline

\end{tabular}
}
\end{table}

\textbf{Steps 4-5:  Risk Assessment and Classification:} 
Steps 4--5 cover risk assessment and classification. In HARA, this is based on severity, exposure, and controllability, resulting in an ASIL classification (see Table~\ref{tab:hara_dnn}). In TARA, risks are analyzed from the attacker's perspective based on impact and feasibility (see Table~\ref{tab:tara_dnn}). To facilitate comparison, we restructure the TARA analysis to begin with DNN limitations and map them to affected assets and threats accordingly. 

Lack of generalization can lead to unsafe behavior in unfamiliar scenarios and also opens attack surfaces through exploitable input gaps, linking both risk domains. This connection is even stronger for lack of robustness, where the same sensitivity to small perturbations causes safety-critical failures and enables adversarial manipulation. Efficiency limitations pose timing-related risks: they affect real-time performance for safety and can be targeted by attackers to overload or delay perception, creating parallel vulnerabilities. While not directly hazardous, lack of explainability undermines traceability, limiting safety validation and facilitating stealthy attacks that avoid detection. Finally, lack of plausibility has a relatively low safety impact. Still, it presents a higher security risk, as attackers may inject semantically invalid yet perceptually plausible inputs that mislead the system without triggering standard safety responses.

\textbf{Step 6: Mitigation:} Based on the derived ASIL in HARA, we define the \textbf{safety goals}, i.e., top-level safety requirements (see Table~\ref{tab:hara_dnn}). The derived safety goals specifically target the mitigation of key hazards posed by DNN limitations in perception, such as ensuring detection reliability in unseen conditions (generalization), maintaining plausible and timely outputs (plausibility and efficiency), enabling traceable failure analysis (explainability), and defending against mali-
cious input manipulation (lack of robustness).

In TARA, a \textbf{risk treatment} decision is determined based on the risk value for each threat scenario. For our use case (see Table~\ref{tab:tara_dnn}), high-risk threat scenarios, such as adversarial attacks on sensor data, perception model outputs, and DNN behavior, require strong mitigation measures, including adversarial training, sensor fusion, or runtime detection. Medium-risk scenarios, e.g., due to explainability gaps or latency, suggest further analysis or partial mitigation, and low-risk items may be accepted or monitored without immediate intervention. Compared to HARA, the TARA outcomes emphasize cybersecurity risk treatment decisions (e.g., monitoring, hardening, isolation) to protect the same assets from intentional exploits.

\section{Conclusion}
This work assessed safety and security risks in AD perception caused by DNN limitations. We reviewed how these limitations are treated in automotive standards, proposed a combined HARA-TARA workflow, and found that their criticality varies depending on whether safety or security is the focus. 

The key limitation of the presented risk analysis is its inherently subjective assignment of risk levels, which were based on the authors' judgment in this work rather than empirical incident data or multi-expert validation, which could be included in future work. The analysis was thus deliberately qualitative and system-level in nature, in line with established risk assessment methodologies and the objective of identifying the most critical DNN insufficiencies. Rather than aiming for exhaustive quantification of threats or failure propagation, the focus was on highlighting the high-risk areas that warrant targeted investigation. The scope was restricted to camera-based perception, reflecting its central role in many AD systems and its relevance to current industry practice. The analysis has identified key limitations, such as lack of robustness and generalization. Our findings highlight the interdependent nature of safety and security in AI-based perception and the need for integrated risk assessments that can capture cross-domain failure propagation.

\begin{credits}
\textbf{\ackname} This work was supported by funding from the Topic Engineering Secure Systems of the Helmholtz Association (HGF) and by KASTEL Security Research Labs (46.23.03).

\end{credits}

\newpage
\clearpage
\bibliographystyle{splncs04}
\bibliography{references}
\end{document}